\documentclass[%
 aip,
 amsmath,amssymb,
 reprint,%
]{revtex4-1}

\usepackage{graphicx}
\usepackage{dcolumn}
\usepackage{bm}

\usepackage[utf8]{inputenc}
\usepackage[T1]{fontenc}
\usepackage{mathptmx}
\usepackage{etoolbox}
\usepackage{multirow}
\graphicspath{{figures/}}

\makeatletter
\def\@email#1#2{%
 \endgroup
 \patchcmd{\titleblock@produce}
  {\frontmatter@RRAPformat}
  {\frontmatter@RRAPformat{\produce@RRAP{*#1\href{mailto:#2}{#2}}}\frontmatter@RRAPformat}
  {}{}
}%
\makeatother
\begin{document}

\preprint{AIP/123-QED}

\title{Universal self-scalings in a micro-co-flowing}
	
\author{Z. L. Wang}
\email[]{wng\_zh@i.shu.edu.cn}
\affiliation{Shanghai Key Laboratory of Mechanics in Energy Engineering, Shanghai Institute of Applied Mathematics and Mechanics, School of Mechanics and Engineering Science, Shanghai University, Shanghai 200444, China}

\date{\today}
\begin{abstract}
On hypothesis of self-scaling co-flows in tapered rectanglar PMMA micro-channels for producing mono-dispersed liquid cells, universal scalings through all liquid detaching regimes are found under a self-similarity frame. Pan-dripping and Pan-jetting regimes are calculated clearly to border at the Weber number approximately $1$ and the capillary number approximately $0.28$ by machine learning classification. The sizes, as well as detaching frequencies, of liquid cells in different flow regimes behave at the same manner and submit to the same law, and such highly consistent behaviors breaking through physical barriers among flow regimes have never been reported.
\end{abstract}
\maketitle

Microfluidic technology has been demonstrated as one of the most effective methods for the generation of mono-dispersed microdroplets/microbubbles \cite{Ganan-Calvo:2001, Rosenfeld:2014, Oomen:2016}. The produced uniform liquid cells, ranging from several micrometers to hundreds of micrometers, are idealized items for the research of chemical reaction and mass-transfer processes \cite{Demello:2006, Martinez:2012, Lang:2012} to miniaturize and optimize complex laboratory procedures as well as the development of low cost biological devices and research \cite{Kleinstreuer:2008, Xu:2009, Wang1:2015, Kashid:2007, Abiev:2013, Arsenjuk:2016}. To explore the optimal shape and size of droplet in channel with perfect internal circulations and unitized bulk/surface reactant time, it is of great interest to make accessible and tunable droplet behaviours to ensure proper chemical or mass transfer performances. Unit liquid cells, generated in slug, dripping and thin jet modes, are mostly studied for their uniformity \cite{Cubaud:2008, Zhao:2018}. Pioneering studies have been made on emulsion mechanisms in these droplet procedures \cite{Garstecki:2005, Utada:2005, Utada:2007, Utada:2008, Guerrero:2020}. The quantitative descriptions about the droplet length, drop-detaching frequency, and dripping-jetting transitions are important common issues \cite{Nunes:2013, Zhu:2017, Favreau:2020}. Also many relevant studies have been reported in literatures involving the diversity of channel geometries, such as T-junction and Y-junction, cross-junction and flow-focusing and some other kinds of micro-fluidic devices \cite{Hashimoto:2010, Castro-Hernandez:2011, Wang:2015, Svetlov:2018, Shen:2018}. Overviews of different geometrical approaches can be found in \cite{Nunes:2013, Dang:2013, Zhu:2017, Montanero:2020, Svetlov:2018}, where owing to the inherent complexity of flow and the diversity of geometric structure, even though the mechanisms of generating fluid units are the same and the phenomena obtained in various literatures are similar, the regularities of liquid cell characteristics both in or among regimes are quite different and have few consistency.

In this letter, the author reports a particular and supreme rectanglar co-flow configuration, named as $\Psi$-junction, that can unify the droplet behaviors of different flow regimes and give very concise and universal control expressions in a self-similarity frame.

\begin{figure*}[t!]
\includegraphics[width=\columnwidth]{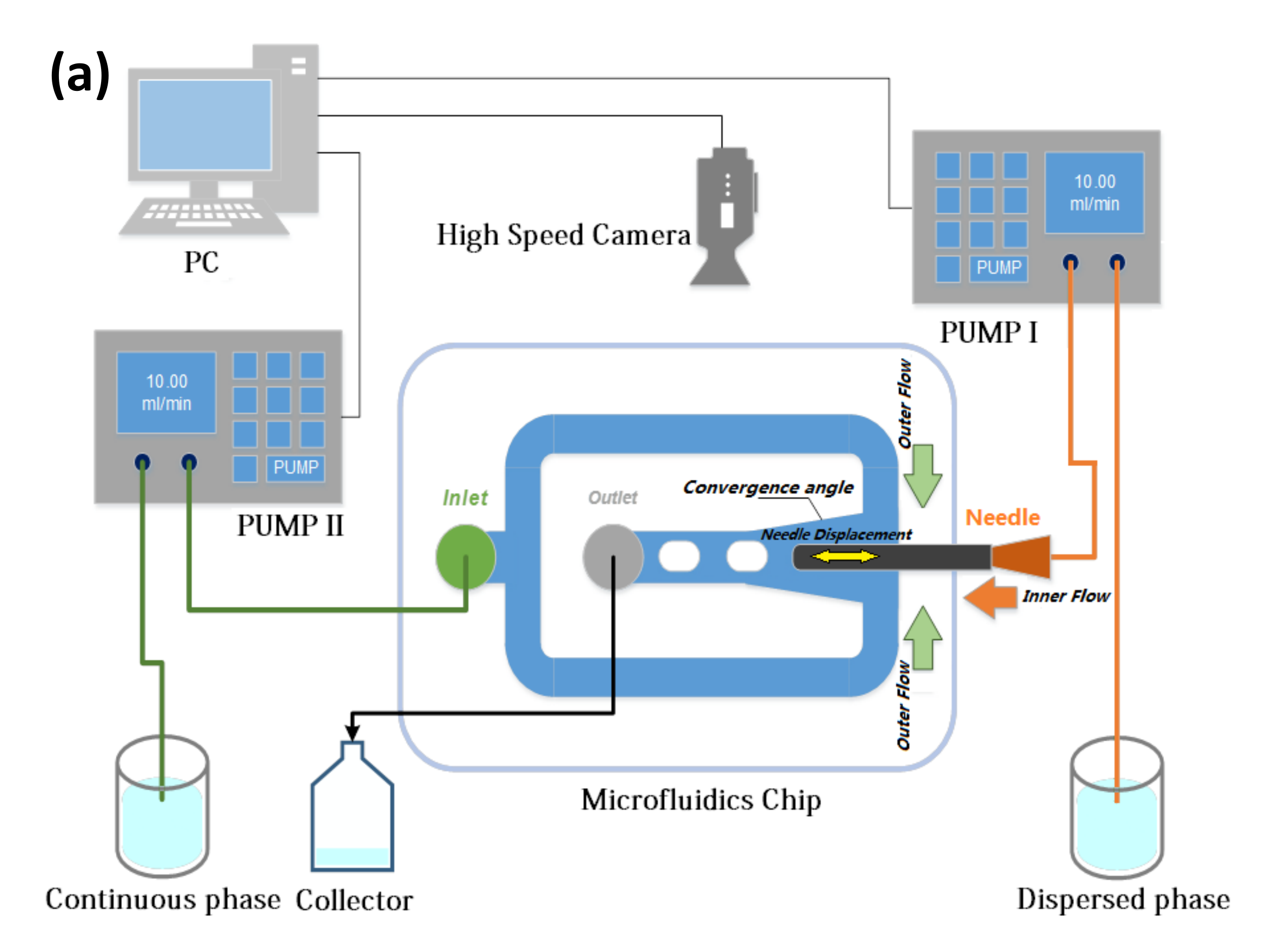}
\includegraphics[width=\columnwidth]{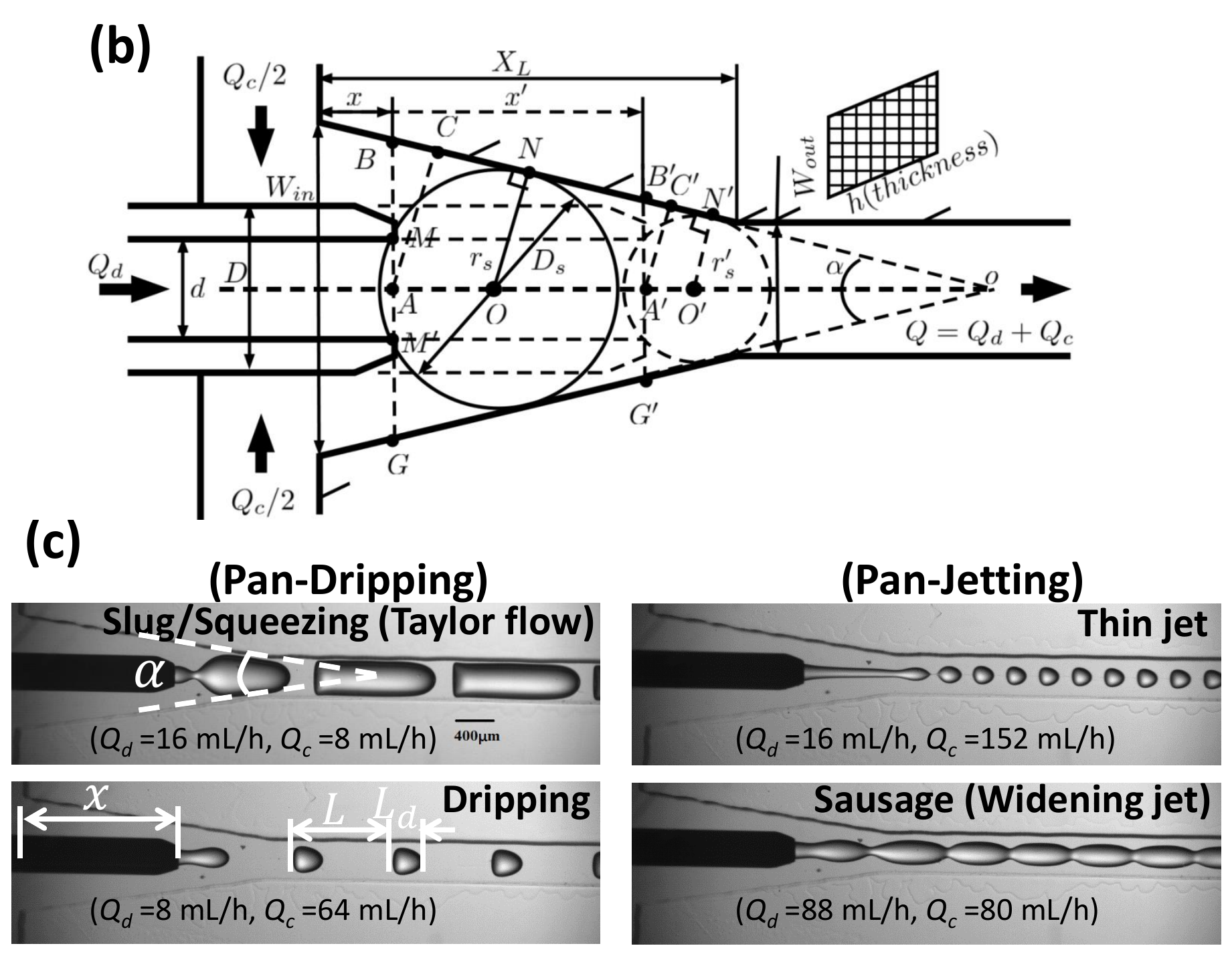}
\caption{Experimental set-up. (a) Schematic of the micro-fluidic device system, (b) self-similar geometry and flow structure of $\Psi$-junction for various needle displacement $x$ and various convergence angel $\alpha$. When $x$ increases to $x'$, and the effective tapered zone changes from $\triangle o B G$ to $\triangle o B' G'$, where point $'o'$ is the intersection point of the two tapered zone extension lines. Two corresponding bulbs represent the largest spherical droplets, whose sizes are obviously limited by the walls, and form in the tapered zone, (c) snapshots of four typical flow patterns in PMMA micro-channels, shown as slug, dripping, thin jet and sausage (widening jet), which are morphologically divided into two groups. The former two are named Pan-dripping and the last two are named Pan-jetting.}
\label{Fig. 1}
\end{figure*}

\paragraph*{Experiments apparatus.}The microchips, fabricated on polymethalmethacrylate (PMMA, over $90\%$ of light transmittance) substrates , are connected with apparatus and image processing tools, as shown in Fig. \ref{Fig. 1}(a). Two micro-injection pumps (LSP02-2A, Lange) supply two liquids. The dispersed phase liquid is inserted into the tapered zone with a flat-end syringe needle. While the continuous phase liquid is pumped into the tapered zone from two by-passes. A high-speed camera (Phantom V611-16G-M, AMETEK) with a micro-lens (AT-X M100 PRO-D, Tokina) images our observations, lighted by a high intensity LED light source.

\paragraph*{Geometry, material properties and parameters.}The $\Psi$-junction has a region can dominate drop-forming process when the two liquids interact, as shown in Fig. \ref{Fig. 1}(b). The cross-sections of channel are rectangular with uniform thickness $h=600\mu m$ (in depth) from side view. The channel width decreases from ${W_{in}} $ to ${W_{out}}$ in a span of ${X_L} = 2680\mu m$ with the convergence angle $\alpha$ defining the tapered zone. The downward width is fixed, i.e., ${W_{out}} = 400\mu m$. Therefore, the upward width, ${W_{in}}=W_{out}+2 X_L $tan$(\alpha/2)$, varies with the convergence angle $\alpha$. The steel syringe needle of inner diameter $d = 200\mu m$ and external diameter $D = 400\mu m$ is inserted into the tapered zone. The needle displacement $x$ starts from the inlet of the tapered zone.

The continuous phase (outer) is lubricating oil (5W-20, from Mobil, USA) and the dispersed phase (inner) is deionized water.  Both the flow rates, ${Q_c}$ and ${Q_d}$, of the two liquids range from $8$ to $152 mL/h$. The interfacial tension between the two phases is $\sigma=20.08 mN/m$, measured with an interfacial tension meter (SL200KS, Kins, USA). The densities of lubricating oil (continuous phase) and deionized water (dispersed phase) are $\rho_c =826 kg/{m^3}$ and $\rho_d =986.2 kg/{m^3}$, respectively. The viscosities of lubricating oil and deionized water are $\mu_c=40.96 mPa \cdot s$ and $\mu_d=1.230 mPa \cdot s$, measured with a rotational rheometer (RS6000, HAAK, Germany). All the material properties are measured at room temperature $ 22^\circ C$, and listed in Table \ref{tab:1}. The capture frequency of high-speed camera is $2000$ to $20000$ frame-per-second according to the period of drop-detaching processes.

\begin{table}[htb]
\caption{Material properties at room temperature $ 22^\circ C$}
    \begin{tabular}{ l | c | c | c }
\hline\noalign{\smallskip}
\hline\noalign{\smallskip}
\multirow{2}*{Two phases} & Viscosity & Density &Surface tension\\
& $\mu (mPa \cdot s)$&$\rho (kg/{m^3})$ & $\sigma (mN/m)$ \\
\hline\noalign{\smallskip}
\hline\noalign{\smallskip}
De-ionized water (D) & 1.230 & 986.2 & \multirow{2}{*}{20.08} \\
Lubricating oil (C) & 40.96 & 826   &\\
\hline
\end{tabular}
\label{tab:1}       
\end{table}

Nine sample PMMA micro-chips, labeled as the $'C'$ group $\{C1$, $C2$, $C3$, $C4$, $C5\}$, the $'3'$ group $\{A3$, $B3$, $C3$, $D3\}$ and sample $E$, are prepared with variations of the convergence angle $\alpha$ and the needle displacement $x$. Sample $E$ is for a straight channel case, which is of the Galilean invariance \cite{Wang:2015} about the needle displacement $x$ owing to the convergence angle $\alpha = 0^\circ$, to give comparisons with results of straight channels \cite{Nunes:2013, Zhu:2017}.

Over a thousand of videos are recorded, and the droplet length ${L_d}$ and the frequency of drop-detaching $f$ are identified by a computer program to ensure data quality. General processes of liquid cell forming \cite{Ahn:2020, Vagner:2021} are omitted here.

Four typical flow patterns , slug, dripping, thin jet and sausage (widening jet) \cite{Cubaud:2008, Nunes:2013}, can be obtained as shown in Fig. \ref{Fig. 1}(c), which are divided into two categories according to the strength of the inertial force relative to the shear force and the surface tension force. The Pan-dripping occurs at smaller inertial forces. The shear stress of the continuous phase in the Pan-jetting regime is large enough to stretch the dispersed phase into a thin jet, the thin jet regime occurs; while the shear stress of dispersed phase is large, the thread of sausage regime happens. The Pan-dripping regime can produce periodic droplets with relatively larger liquid size. In the Pan-jetting regime, smaller drops can be generated only in the thin jet regime, just like that of flow-focusing or circular co-flowing \cite{Chong:2016, Guerrero:2020}.

\begin{figure}[t!]
\includegraphics[width=.95\columnwidth]{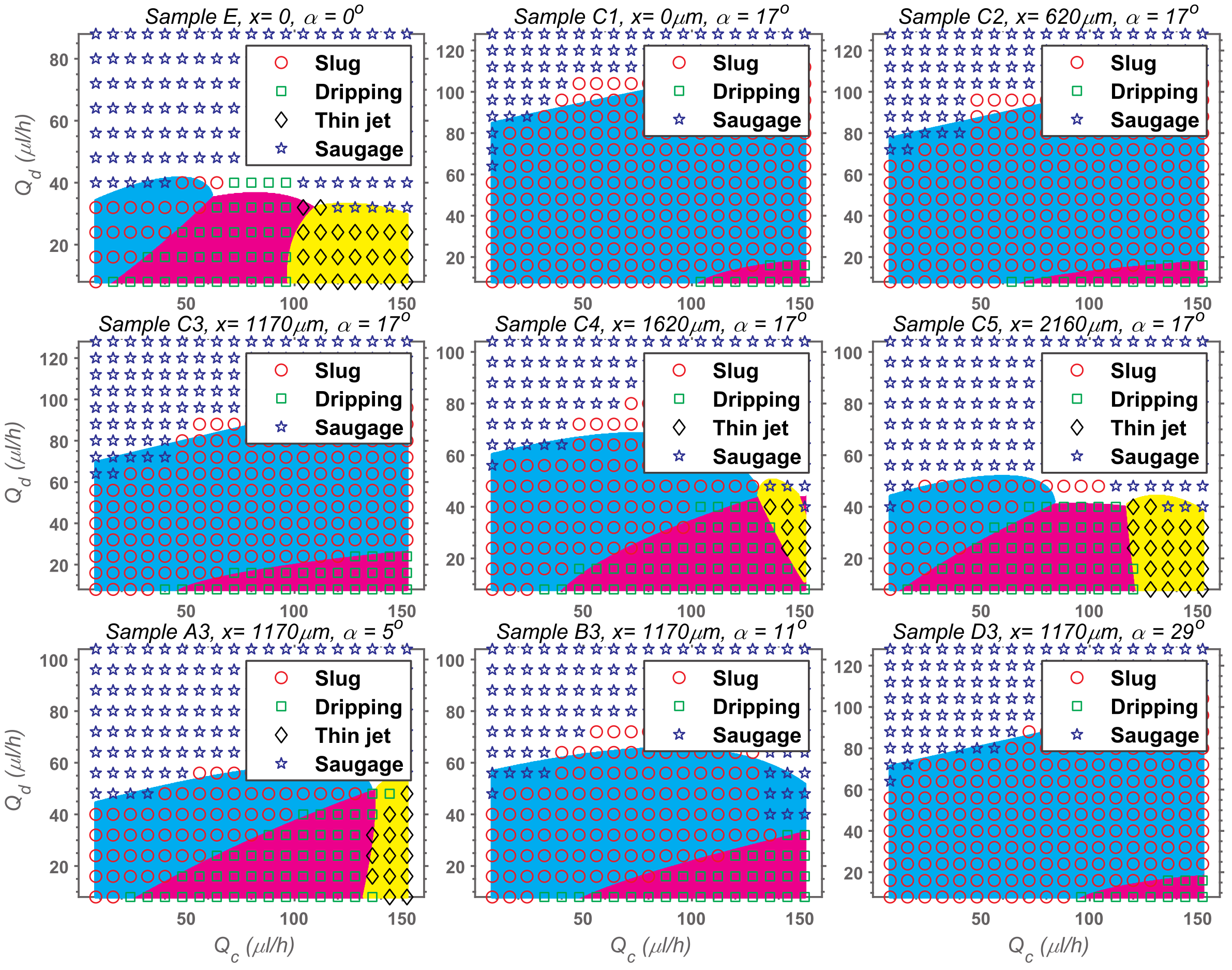}
\caption{Flow maps in $Q_c$-$Q_d$ space ($E$ for straight tube, $'C'$ group $\{C1$, $C2$, $C3$, $C4$, $C5\}$ for $x=\{$0, 620, 1170, 1620, and 2160 $\mu m\}$ at fixed $\alpha=17 ^\circ$, and $'3'$ group $\{A3$, $B3$, $C3$, $D3\}$ for $\alpha=$$\{5^\circ$, $11^\circ$, $17^\circ$, and $29^\circ \}$ at fixed $x=1170 \mu m$). The background colors are calculated by machine learning algorithm using MATLAB ($fitcdiscr$, $Discriminant$ $Analysis$ $Classification$ $Function$).}
\label{Fig. 2}
\end{figure}

The flow maps are shown in Fig. \ref{Fig. 2}. The colored backgrounds are given by the discriminant analysis classification function $fitcdiscr$ in MATLAB, showing the parameter areas covering the corresponding flow patterns. The total nine subgraphs are arranged in a $3 \times 3$ matrix form $\{E$, $C1$, $C2$; $C3$, $C4$, $C5$; $A3$, $B3$, $D3\}$. Sample $E$, a straight tube case, is used as a reference. Sample $C5$ is the most close to sample $E$ to give similar flow maps, where all four flow patterns exist. $'C'$ group show the effect of needle position changes on the flow patterns. With deepening $x$, the regime of the widening jet expands, the slug regime moves toward the origin point of the coordinate, and the dripping and thin jet prevalent regimes gradually appear from the low-right side of the coordinates. Finally, the borders of these three regimes interconnect at $Q_c \approx 100 ml/h$, and meet the widening jet regime at $Q_d \approx 40 ml/h$. The $'3'$ group shows the effects of $\alpha$. The increasing $\alpha$ expands the slug regime, squeezes the other three flow regimes, and gradually kicking the thin jet regime out of the maps. Increasing of $x$ takes contrary effects. It's rational to believe that there is a correlationship between $x$ and $\alpha$ on flow behaviours.

\paragraph*{Non-dimensionalization under a self-similarity frame.}It is important to scale the tapered zone up to reflect the variations of $x$ and $\alpha$ on drop-producing processes. The usual $W_{out}$ as a fixed value is excluded at first. In Fig. \ref{Fig. 1}(b), line segment $\overline{BG}$, $\overline{BA}$, $\overline{CA}$, or $\overline{NO}$ could be selective to give the information of the drop-forming position. However, these line segments lack adaptability to the variation of $\alpha$. By observations, it can be assumed that all geometric changes and physical phenomenon changes in the tapered region are self-scaling processes. To give a dimensional feature with uniformity and symmetry by taking self-similarities into account, we define a local virtual width $W_{Local}$, which consists of two parts as
\begin{equation}
W_{Local} \sim \widetilde{L} \Re(\triangle),
\end{equation}
where $\widetilde{L}$ represents a selected length feature, and $\Re(\triangle)$ is a shape factor. See Fig. \ref{Fig. 1}(b), we choose $\widetilde{L} = \overline{oA}$, which varies with both the changes of $x$ and $\alpha$. Point $'o'$ is the center of fan area enclosing the tapered region. Hence, $\overline{oA}$ is the distance from the cross-section of needle exit to the center of gyration of the fan, and $'o'$ is nearly fixed with varying of $x$ or $\alpha$. When $x$ moves to $x'$, the extended tapered zone becomes smaller from  $\triangle BGo$ to $\triangle B'G'o$. Then the solid bulb turns into the dotted bulb, and both circles are the maximum spherical dispersed drops with radii $r_s$ and $r'_s$. Pairs of similar triangles, $\triangle BGo \cong \triangle B'G'o$, $\triangle BAo \cong \triangle B'A'o$ and $\triangle CAo \cong \triangle C'A'o$, form in this process. It should also be noted that all the triangles are similar. Hence, the shape factor $\Re(\triangle)$ is a perfect form for describing these clusters of triangles. It is selected as $\Re(\triangle)=\overline{CA}/(\overline{CA}+\overline{oA})$. Here, $\overline{CA}$ and $\overline{oA}$ are the two sides of a right triangle $\triangle CAo$. $\overline{CA}$ is also the shortest distance from the needle exit to the tapered borders. Therefore, the virtual scale of local width can be expressed by
\begin{equation}
W_{Local}=2 {\overline{oA}} \cdot \overline{CA}/(\overline{oA}+\overline{CA}).
\end{equation}
Intrinsically, $W_{Local}$ should be a width dimension, however, it maintains symmetry for axial and transverse directions here. Furthermore, when $x$ moves to $x'$, the flow-rate-ratio $Q_d/Q_c$ keeps constant, the superficial velocity ratio between two cross-sections should be proportional to ${W_{Local}}^{-1}$ according to kinematical self-similarity. We can write down the superficial velocity of the dispersed phase as
\begin{equation}
u_d = {Q_d/(\pi \cdot d^2/4)} \cdot {(W_{local}/W_{out})}^{-1},
\end{equation}
where $W_{local}$ is divided by $W_{out}$ for normalization. $Q_d/(\pi \cdot d^2/4)$ is the flow rate over area of needle cross-section, which is the normal definition of superficial velocity.

While the superficial velocity of the continuous phase takes form of
\begin{equation}
u_c = Q_c/((W_{local} \cdot h)-Q_d/u_d),
\end{equation}
to maintain mass conservation. Here, $W_{local} \cdot h$ represents the area of total local cross-section, and $Q_d/u_d$ means the area occupied by the dispersed phase.

Then, with these length and velocities, we can build up our non-dimensional groups.
The capillary number of the continuous phase is expressed as
\begin{equation}
Ca_{c} = {\mu_c}{u_c}/\sigma,
\end{equation}
which represents the ratio of the viscous force to the interfacial tension force of the continuous phase. The Weber number of the dispersed phase is
\begin{equation}
We_{d} ={\rho_d}{u_d}^2{d}/\sigma,
\end{equation}
which represents the ratio of the inertial force to the interfacial tension force of the dispersed phase.
Also, we define another capillary number for both liquids, which can be described as
\begin{equation}
Ca_{TP} ={\mu_c \cdot U_{TP}}/\sigma,
\end{equation}
where $U_{TP}=(Q_d+Q_c)/(W_{local} \cdot h)$ is the total superficial velocity of both phases defined at local cross-sections.
Other dimensionless numbers, such as Reynolds number and Bond number, do not affect our expressions here.

\begin{figure}[t!]
\includegraphics[width=.95\columnwidth]{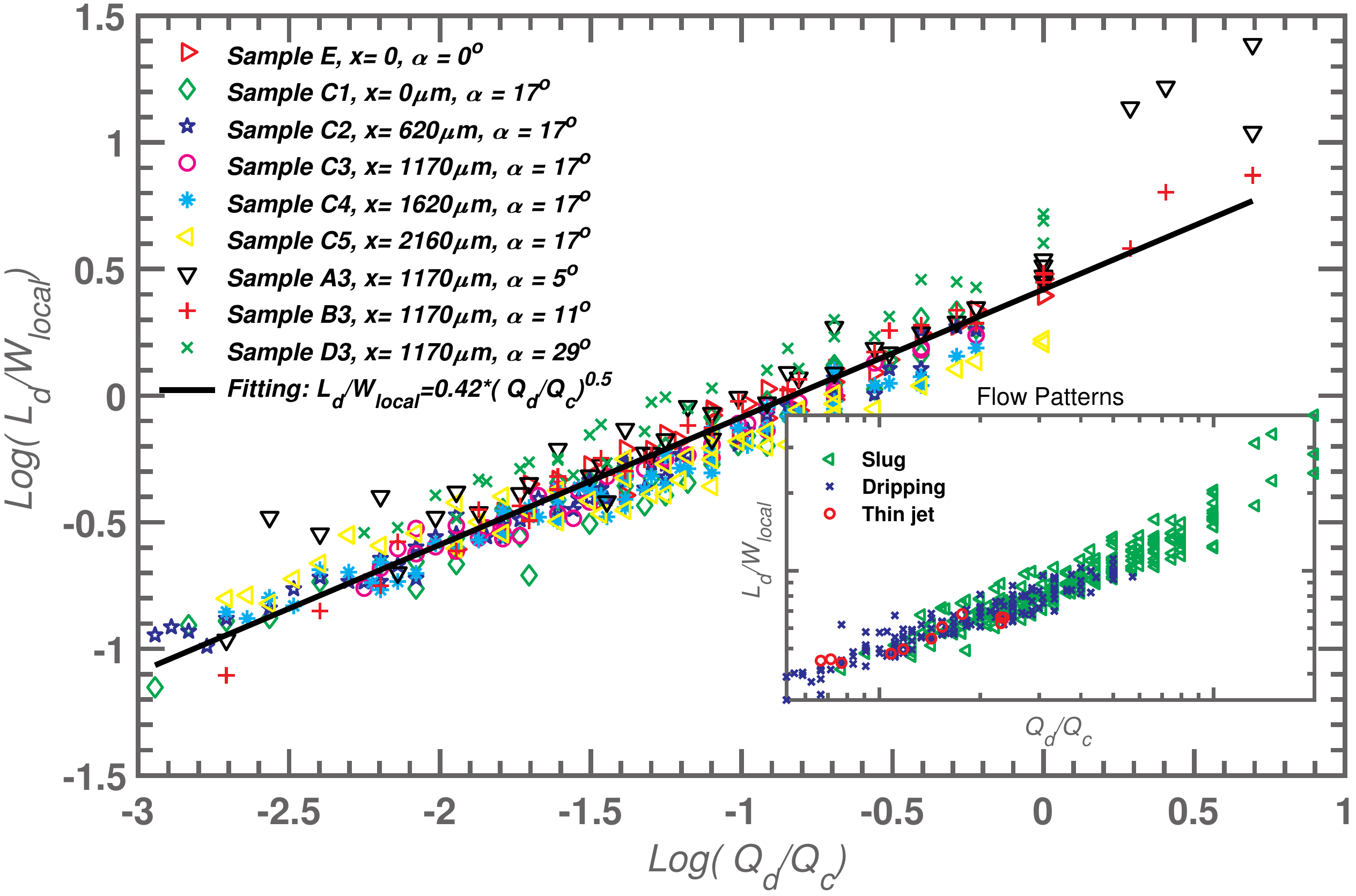}
\caption{Log-log plot about droplet length ${L_d}/{W_{Local}}$ varying with the flow-rate-ratio ${Q_d}/{Q_c}$ for all the nine samples. The slope of the solid line is close to $1/2$. Inset: Data from different flow patterns (slug ($\triangleleft$), dripping ($\times$), and thin jet ($\circ$)).}
\label{Fig. 3}
\end{figure}

\paragraph*{Universal scalings.}By using the self-similarity characteristic quantities, the droplet length is plotted in Fig. \ref{Fig. 3}. The data covers all drop-detaching regimes. The developed correlation is drawn into a solid line laying through the data symbols to give an empirical relation using the least squares regression analysis, and described as
\begin{equation}
L_d / W_{Local} = 0.42 (Q_d/Q_c)^{0.5}.
\label{equ:dimen}
\end{equation}
Eq. \ref{equ:dimen} is comparable to $L_d \sim (Q_d/Q_c)^{1/2}$, well-known for the thin jet as a $'$Flow-Rate-Controlled-Flow$'$  \cite{Cubaud:2008, Sheu:2010}. Hence, the present scaling extends the $'$Flow-Rate-Controlled-Flow$'$ from the thin jet regime into all the drop-detaching regimes. 

\begin{figure}[t!]
\includegraphics[width=.95\columnwidth]{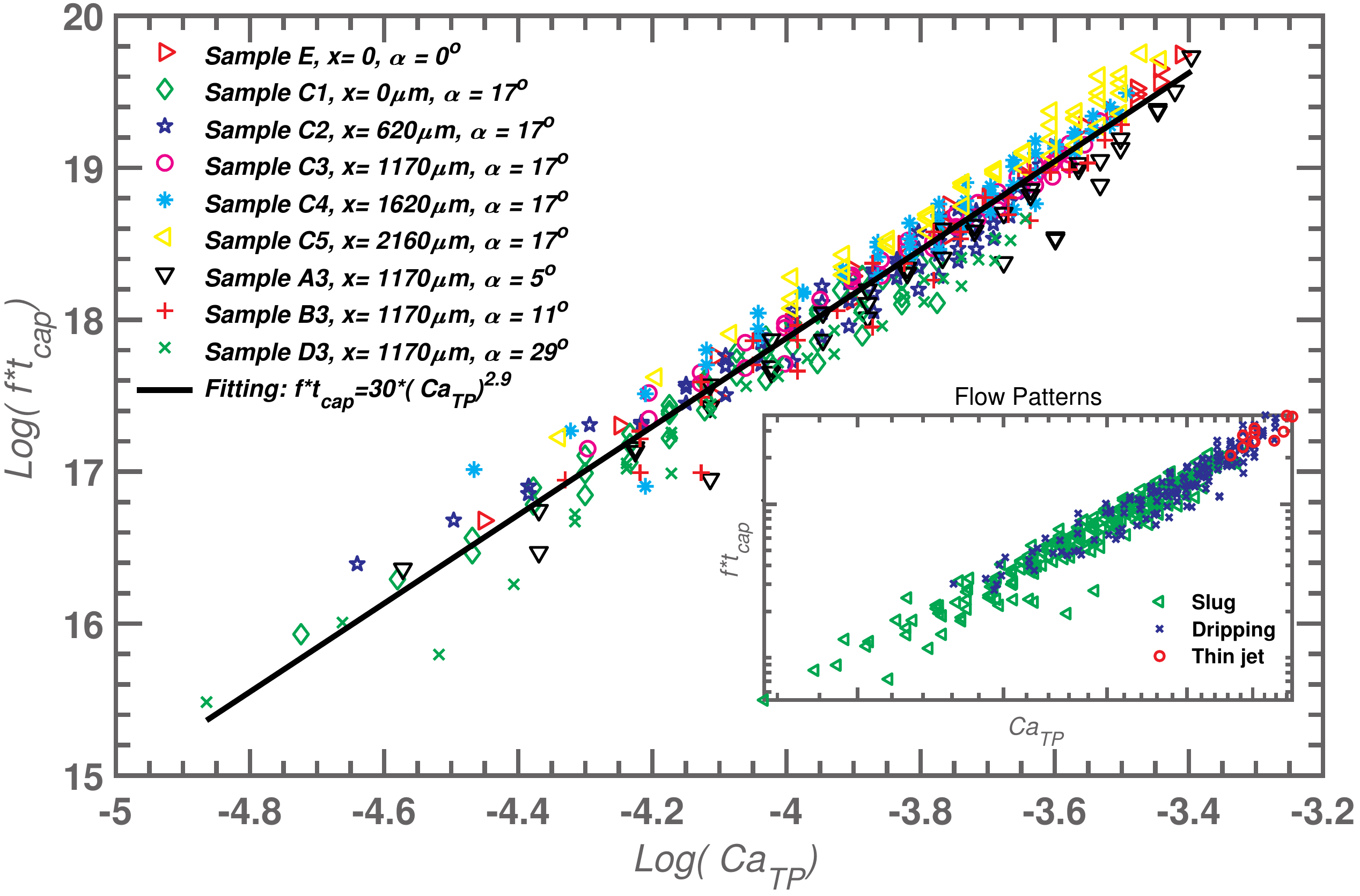}
\caption{Log-log plot about non-dimensionalized frequency $f\cdot t_{cap}$ varying with the two-phase capillary number $Ca_{TP}$ for all the nine samples. The slope of the solid line is close to $3$. Inset: Data from different flow patterns (slug ($\triangleleft$), dripping ($\times$), and thin jet ($\circ$)).}
\label{Fig. 4}
\end{figure}

The drop-detaching frequency is also drawn in Fig. \ref{Fig. 4}, and expressed as
\begin{equation}
f\cdot t_{cap} = 30 {Ca_{TP}}^{2.9},
\label{Eq:freq}
\end{equation}
where the capillary time $t_{cap} = \sqrt{\rho_d  \cdot d^3/\sigma}$. The dripping mode has similar formula $f\cdot t_{cap} = 0.79 {Ca_c}^{1.35}$ in Liu(2018) \cite{Liu:2018}. Scaling in Eq. \ref{Eq:freq} also extends its effectiveness from the dripping regime into all the drop-detaching regimes. Moreover, Eq. 3 in Guerrero(2020) \cite{Guerrero:2020} gave another formula about the periodic formation of droplet at the maximum frequency of drop-detaching for the dripping regime as follows:
\begin{equation}
f^{max} \sim \sqrt{1/(\rho_i \gamma ^5 d_{tip} ^3)} {\mu_o U_o}^3,
\label{Eq:Guerrero}
\end{equation}
where $d_{tip}$ is the diameter of the dispersed liquid tip, $\rho_i$ is its density, $\mu_o$ is the viscosity of the outer medium, $U_o$ is the speed of the outer medium, and $\gamma$ is the interfacial tension. Note that the exponential in Guerrero(2020) \cite{Guerrero:2020} over $U_o$ is 3 and very close to the present exponential over $Ca_{TP}$, 2.9. By rounding off 2.9 to 3, the correlation Eq. \ref{Eq:freq} is translated into the primitive form as
\begin{equation}
f \sim Ca_{TP}^{3} \cdot t_{cap} \sim \sqrt{1/(\rho_d \sigma^5 d^3)} {\mu_c U_{TP}}^3.
\label{Eq:fme}
\end{equation}
Eq. \ref{Eq:fme} and Eq. \ref{Eq:Guerrero} share the same form, while the present formula Eq. \ref{Eq:fme} covers all drop-detaching regimes under a self-similarity meanings.

\begin{figure}[!htb]\centering
\includegraphics[width=\linewidth]{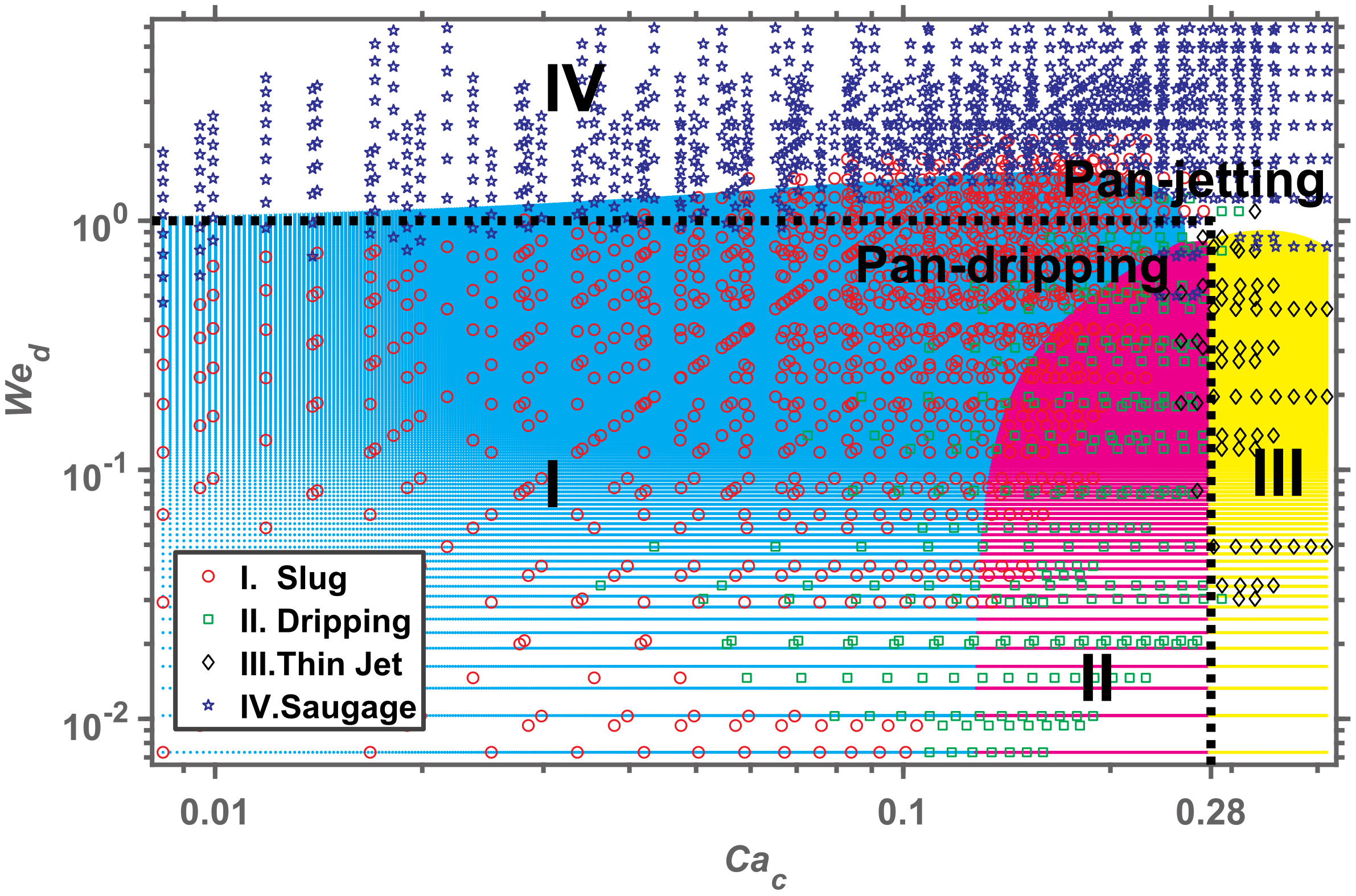}
\caption{Flow pattern map in space of $We_d$ and $Ca_c$ is divided into two parts: Pan-dripping (slug ($\circ$), dripping ($\Box$)) and Pan-jetting (thin jet ($\Diamond$), sausage ($\star$)). The transition-lines are plotted according to the background colors, which are calculated by machine learning of all the data using MATLAB ($fitcdiscr$). These results are very consistent, graphically and quantitatively, with the plots in literatures \cite{Utada:2007, Meyer:2009, Mak:2017, Guerrero:2020}, which are for circular co-flows.}
\label{Fig. 5}
\end{figure}

Furthermore, it can be found that the flow map for Pan-dripping and Pan-jetting regimes is determined by two dimensionless quantities: the continuous phase capillary number $Ca_c$ and the dispersed phase Weber number $We_d$. These two numbers reflect the competitions between viscous force or inertial force, respectively, against surface tension force. The transition from Pan-dripping to Pan-jetting seems to occur when $We_d \approx 1 $ from Pan-dripping to widening jet and $Ca_c \approx 0.28$ from Pan-dripping to thin jet (see Fig. \ref{Fig. 5}). These critical values are very close to those of circular co-flows \cite{Utada:2007, Mak:2017, Guerrero:2020, Deng:2017}.

\paragraph*{Summary and conclusion.}Digital micro-fluidics attempts to find perfect ideas of producing mono-dispersed fluid cells, it may have discovered one here. Self-scaling processes are revealed and a self-similarity frame is build up to derive universal scalings through all liquid detaching regimes in tapered rectanglar PMMA micro-channels. Pan-dripping and Pan-jetting regimes, distinguished by machine learning classification, are found clearly bordered at $We_d \approx 1$ and $Ca_c \approx 0.28$ under the self-similarity frame. The liquid detaching regimes, including slug, dripping, and thin jet, scale their drop lengths or detaching frequencies at the same manners and submit to the same laws, though their physical mechanisms are far differentiated. Specifically, the droplet length is proportional to the square-root of flow-rate-ratio, while the drop generation frequency is proportional to the cubic power of two-phase capillary number $Ca_{TP}$ without scruple of flow patterns. Such highly consistent behaviors through physical mechanisms have never been reported. The most important task of the efficiency of mini and micro devices is to extend the maintenance of the regular Taylor flow (slug) regime to make uniformed fluid cells. Therefore, this work is of great advantages.

\begin{acknowledgments}
This research is sponsored by National Key Research and Development Program (2017YFB0404503) and National Science Foundation of China (No.11172163). The author would like to express his thanks for the helpful discussion with Prof. Z.W. Zhou, Prof. G.H. Hu, Prof. J.H. Zhang, and Prof. J.S. Zhang.
\end{acknowledgments}

\section*{Conflict of Interest}
The author declares no conflict of interest to this work.
\section*{DATA AVAILABILITY}
The data that support the findings of this study are available
from the corresponding author upon reasonable request.

\bibliographystyle{apsrev4-1}
\bibliography{universal}

\end{document}